\documentclass{svjour2}
 \journalname{JSP}

\begin{document}


\title{Thermodynamic restrictions on \\ statistics
of molecular random walks}

\titlerunning{On statistics of molecular random walks}

\author{Yuriy E. Kuzovlev}

\institute{Donetsk Institute for Physics and Technology of NASU\\
ul.\,\,R.\,Luxemburg 72,\, 83114 Donetsk, Ukraine\\
\email{kuzovlev@kinetic.ac.donetsk.ua} }


\date{}
\maketitle

\begin{abstract}
It is shown that time reversibility of Hamiltonian microscopic
dynamics and Gibbs canonical statistical ensemble of initial
conditions for it together produce an exact virial expansion for
probability distribution of path of molecular Brownian particle in a
fluid. This expansion leads to inequality connecting logarithmic
derivative of the distribution with respect to fluid density and
characteristic volume occupied by pair statistical correlation
between the path and fluid molecules. Due to the inequality,
finiteness of this volume means that asymptotic of the distribution
is essentially non-Gaussian. For principal case when fluid is dilute
gas it is demonstrated that the pair correlation volume is actually
finite and bounded above. Therefore even under the Boltzmann-Grad
limit the path distribution possess power-law long tails (cut off at
distances of ballistic flight).


\end{abstract}







\pagenumbering{arabic}

\section{Introduction}

Thermal motion of molecules in fluids or quasi-particles in solids
gives us examples of chaotic walk of small molecular-size ``Brownian
particle''. As a rule, considerations of such walk, or ``molecular
Brownian motion'', involve more or less amount of stochasticity
constantly brought in from random ``thermal bath'' or random choices
like coin tossing, etc. Many understand that this way of thinking
conflicts with principles of statistical mechanics where the only
random choice is that of initial conditions for Hamilton equations.
Some understand that even the Boltzmann's ``Stosszahlansatz'' and
thus validity of the Boltzmann equation, even for dilute gas under
the Boltzmann-Grad limit, remain unproved \cite{re,dor}. All the more
stochastic probability-theoretical models of molecular motion in fact
stay based on faith rather than Hamiltonian dynamics.

Such state of affairs was predicted by Krylov in \cite{kr} (firstly
this book was published in 1950 in Russian). According to him, actual
(cause-and-effect) independence of events in reality, on concrete
phase trajectory of a system under interest, generally does not imply
statistical independence of same events in theory, in statistical
ensemble of trajectories induced by a given ensemble of their initial
conditions. For instance, ensemble-average density of pair collisions
of gas atoms, along with pair distribution function of colliding
atoms, in spatially non-uniform gas does not reduce, even under the
Boltzmann-Grad limit, to (quadratic form of) one-particle
distribution function \cite{i1}.

In such case kinetic theory loses its traditionally most powerful
tool. To compensate this loss, one can examine other tools, e.g.
exact ``generalized fluctuation-dissipation relations'' (FDR)
\cite{j1,j2,p} implied by determinism and time reversibility of
microscopic dynamics. In the present paper we exploit one of simplest
FDR what correspond to the classical Gibbs canonical
thermodynamically equilibrium ensemble.

Main object of our interest is probability distribution,
$\,V_0(t,\Delta{\bf R})\,$, of displacement, $\,\Delta {\bf R}\,$, of
molecular Brownian particle (BP) during time interval $\,(0,t)\,$
much longer than its mean free-flight time. We transform the FDR into
an original virial expansion of $\,V_0(t,\Delta{\bf R})\,$ connecting
its $\,n$-order derivatives with respect to density of gas (or,
generally, fluid) and $\,(n+1)$-particle probability distribution
functions and correlation functions describing statistical
correlations between the BP's displacement and gas particles. As
combined with positiveness of distribution functions, the virial
expansion produces differential inequalities to be satisfied by
$\,V_0(t,\Delta{\bf R})\,$. We show that already first of them
imposes a limitation on possible steepness of $\,V_0(t,\Delta{\bf
R})$'s tails at $\,|{\bf R}| > \sqrt{6Dt}\,$, where $\,D\,$ is BP's
diffusivity, certainly forbidding exponentially vanishing tails. In
particular, the Gaussian asymptotic is forbidden, first of all, for
the Boltzmann-Grad gas.

\section{Fluctuation-dissipation relations}

Let us consider a molecular-size ``Brownian particle'' (BP) among
$\,N\gg 1\,$ atoms in a box with volume $\,\Omega \,$ under the
thermodynamical limit: $N\rightarrow\infty \,$,
$\,\Omega\rightarrow\infty \,$, $\,N/\Omega =\nu_{\,0}  =
\,$\,\,const\,. Introduce designations $\,{\bf q}=\{{\bf R},{\bf
r}_1,...\,,{\bf r}_N\}$ and $\,{\bf p}=\{{\bf P},{\bf p}_1,...\,,{\bf
p}_N\}\,$ for canonical coordinates and momenta of our system,
respectively, and
\[
H({\bf q},{\bf p})\,=\, \frac {{\bf P}^2}{2M}+\sum_{j} \frac {{\bf
p}_j^2}{2m}\,+\,\sum_{j}U_{ab}(|{\bf r}_j-{\bf
R}|)\,+\,\,\,\,\,\,\,\,\,\,\,\,\,\,\,\,\,\,\,\,\,
\,\,\,\,\,\,\,\,\,\,\,\,
\]
\[
\,\,\,\,\,\,\,\,\,\,\,\,\,\,\,\,\,\,\,\,\,\,\,\,
\,\,\,\,\,\,\,\,+\,\sum_{j<k}U_{aa}(|{\bf r}_j-{\bf
r}_k|)\,+\,U_{box}({\bf R})+\sum_{j} U_{box}({\bf r}_j)
\]
for its Hamiltonian in absence of external fields, so that $\,H({\bf
q},-{\bf p})=H({\bf q},{\bf p})\,$. Besides, let $\,{\bf q}(t)={\bf
q}(t,{\bf q},{\bf p})\,$ and $\,{\bf p}(t)={\bf p}(t,{\bf q},{\bf
p})\,$ be solution to corresponding Hamilton equations under initial
conditions $\,{\bf q}(0)={\bf q}\,$ and $\,{\bf p}(0)={\bf p}\,$, and
$\,\rho_{\,eq}({\bf q},{\bf p})\propto \exp{[-H({\bf q},{\bf
p})/T\,]}\,$ equilibrium probability distribution for these
conditions (normalizing factor is omitted). Then for any suitable
functions $\,A({\bf q},{\bf p})\,$ and $\,B({\bf q},{\bf p})\,$ we
can write
\begin{eqnarray}
\int\int \,A({\bf q}(t),{\bf p}(t))\,B({\bf q},{\bf
p})\,\,\rho_{\,eq}({\bf q},{\bf p})\,d{\bf q}\,d{\bf p}\,=
\,\,\,\,\,\,\, \,\,\,\,\,\,\,\,\,\,\,\,\,\,\,\,\,
\,\,\,\,\,\,\,\,\,\,\label{sim}
\\ \,\,\,\,\,\,\,\,\,\,\,\,\,\,\,\,\,\,\,\,\,\,
\,\,\,\,\,\,\,\,\,\,\,\,\, \,\,\,\,\,\,\,\,\,\,\,\,\,\,=\,\int\int
\,B({\bf q}(t),-{\bf p}(t))\,A({\bf q},-{\bf p})\,\,\rho_{\,eq}({\bf
q},{\bf p})\,d{\bf q}\,d{\bf p}\,\nonumber
\end{eqnarray}
This quite obvious identity is very particular form of the mentioned
FDR. But we want to concretize it even greater by choosing
\begin{equation}
\begin{array}{l}
A({\bf q},{\bf p})\,=\,\delta({\bf R}-{\bf R}^{\prime})\,\,\,,\\
B({\bf q},{\bf p})\,=\,\Omega\,\,\delta({\bf R}-{\bf
R}_0)\,\delta({\bf P}-{\bf P}_0)\exp{[\,-\sum_{j} U({\bf r}_j,{\bf
p}_j)/T\,]}\,\\\,\,\,\,\,\,\,\,\,\,\,\,\,\,
\,\,\,\,\,\,\,\,=\,\Omega\,\,\delta({\bf R}-{\bf R}_0)\,\delta({\bf
P}-{\bf P}_0)\prod_{j}\, [\,1+\psi({\bf r}_j,{\bf p}_j)\,]
\,\,\,,\label{ch}
\end{array}
\end{equation}
where\,\, $\,\psi({\bf r},{\bf p})\,\equiv\,\exp{[-\,U({\bf r},{\bf
p})/T\,]}-1\,$.

We assume that the potentials of atom-atom interaction,
$\,U_{aa}(r)\,$, and BP-atom interaction, $\,U_{ab}(r)\,$, satisfy
conventional standards of kinetic theory of fluids \cite{re,bog,uf}
which allow to speak about solutions of the Hamilton equations at
arbitrary large $\,N\,$ at least in statistical language, in terms of
the distribution functions (DF) as introduced by Bogolyubov
\cite{bog}. What is for the function $\,U({\bf r},{\bf p})\,$, or
$\,\psi({\bf r},{\bf p})\,$, it is good enough if
\begin{equation}
\begin{array}{l}
\int |\phi({\bf r})|\,d{\bf r}\,
<\,\infty\,\,\,,\,\,\,\,\,\,\,\phi({\bf r})\,\equiv\,\int \psi({\bf
r},{\bf p})\,G_m({\bf p})\,d{\bf p}\,\,\,, \label{cond}
\end{array}
\end{equation}
with\,\, $\,G_m({\bf p})=(2\pi Tm)^{-3/2}\exp{(-{\bf p}^2/2Tm)}\,$\,
being equilibrium Maxwell momentum distribution. At these assumptions
the thermodynamical limit of (\ref{sim}), when the box walls move
away to infinity, can be treated in full analogy with \cite{bog}.
Firstly, consider right-hand side of (\ref{sim}).

\section{Many-particle distribution functions \\
and correlation dressing of BP}

The right-hand side of (\ref{sim}), after substitution of (\ref{ch}),
tends to
\[
\begin{array}{l}
\int\int B({\bf q}(t),-{\bf p}(t))\,A({\bf q},-{\bf
p})\,\rho_{\,eq}({\bf q},{\bf p})\,d{\bf q}\,d{\bf
p}\,=\mathcal{F}\{t,{\bf R}_0,-{\bf P}_0,\psi\,|{\bf
R}^{\prime}\}\,\,,
\end{array}
\]
\begin{eqnarray}
\mathcal{F}\{t,{\bf R}_0,{\bf P}_0,\,\psi\,|{\bf
R}^{\prime}\}\,\equiv \,V_0(t,{\bf R}_0,{\bf P}_0|{\bf
R}^{\prime})\,+\,\,\,\,\,\, \,\,\,\,\,\,\,\,\,\,\,\,\,\,\,\,\,
\,\,\,\,\,\,\label{gf} \\
+\sum_{n\,=1}^{\infty } \frac {\nu_{\,0}^n}{n!}\int^n_{r\times p}
F_n(t,{\bf R}_0, {\bf r}_1\,...\,{\bf r}_n,{\bf P}_0,{\bf
p}_1\,...\,{\bf p}_n|{\bf R}^{\prime})\prod_{j\,=1}^n \psi({\bf
r}_j,-{\bf p}_j)\,\,\,,\nonumber
\end{eqnarray}
where $\,\int^n_{r\times p}\,$ means integration over $\,n\,$
coordinates $\,{\bf r}_1\,...\,{\bf r}_n\,$ and momenta $\,{\bf
p}_1\,...\,{\bf p}_n\,$\,;\, function\, $\,V_0(t,{\bf R}_0,{\bf
P}_0|{\bf R}^{\prime})\,$ is conditional probability density of
finding BP at $\,t>0\,$ at point $\,{\bf R}_0\,$\, with momentum
$\,{\bf P}_0\,$ under condition that BP had started at $\,t=0\,$ from
point ${\bf R}^{\prime}\,$\,;\, function $\,F_n\,$\, is joint
conditional probability density of this event and simultaneously
finding some $\,n\,$ atoms at points $\,{\bf r}_j\,$\, with momentums
$\,{\bf p}_j\,$ under the same condition.

In respect to atoms, all these $\,F_n\,$ are completely analogous to
usual non-normalized DF of infinite gas \cite{bog}. The role of
normalization is played by uncoupling of inter-particle correlations
at infinity:
\begin{equation}
\begin{array}{l}
F_n(...\,{\bf r}_k\,...\,{\bf
p}_k\,...\,)\,\rightarrow\,F_{n-1}(...\,{\bf r}_{k-1}, {\bf
r}_{k+1}...\,\,{\bf
p}_{k-1}, {\bf p}_{k+1}...\,)\,G_m({\bf p}_k)\,\,\,,\\
F_1(t,{\bf R}_0,{\bf P}_0,{\bf r}_1,{\bf p}_1 |{\bf
R}^{\prime})\,\rightarrow\, V_0(t,{\bf R}_0,{\bf P}_0|{\bf
R}^{\prime}\,)\,G_m({\bf p}_1)\,\,\,,\label{unc}
\end{array}
\end{equation}
when $\,{\bf r}_k\rightarrow \infty\,$ and $\,{\bf r}_1\rightarrow
\infty\,$, respectively. But in respect to BP all DF from (\ref{gf})
are normalized in literal sense. In particular,
\[
\begin{array}{l}
\int V_0(t,{\bf R}_0,{\bf P}_0|{\bf R}^{\prime})\,d{\bf P}_0
\,=\,V_0(t,{\bf R}_0-{\bf R}^{\prime})\,\,\,,\,\,\,\,
 \int V_0(t,{\bf R}_0-{\bf R}^{\prime})\,d{\bf R}_0\, =\,1\,
\end{array}
\]
In respect to $\,\psi\,$, expression $\,\mathcal{F}\{t,{\bf R}_0,{\bf
P}_0,\,\psi\,|{\bf R}^{\prime}\}\,$ represents generating functional
of the DF quite similar to the Bogolyubov's one \cite{bog}. Under
condition (\ref{cond}) the series in (\ref{gf}) converges, hence,
this functional is well defined.

From (\ref{gf}) it is obvious that initial conditions to these DF
appear as
\[
\begin{array}{l}
V_0(0,{\bf R}_0,{\bf P}_0|{\bf R}^{\prime})\,=\,\delta({\bf R}_0-{\bf
R}^{\prime})\,G_M({\bf P}_0) \,\,\,,\,\,\,\,\,\,\,\,
\end{array}
\]
\[
\begin{array}{l}
F_n(0,{\bf R}_0, {\bf r}_1\,...\,{\bf r}_n,{\bf P}_0,{\bf
p}_1\,...\,{\bf p}_n|{\bf R}^{\prime})\,=\,
\,\,\,\,\,\,\,\,\,\,\,\,\,\,\,\,\,\,\,\,\,\,\,\, \\
\,\,\,\,\,\,\,\,\,\,\,\,\,\,\,\,\,\,\,\,\,\,\,\,
\,\,\,\,\,\,\,\,\,\,\,\,=\,F_n^{(eq)}({\bf r}_1...\,{\bf r}_n|{\bf
R}_0)\,\delta({\bf R}_0-{\bf R}^{\prime})\,G_M({\bf
P}_0)\prod_{j\,=1}^n G_m({\bf p}_j)\,\,\,, 
\end{array}
\]
\[
\begin{array}{c}
\mathcal{F}\{0,\,{\bf R}_0,{\bf P}_0,\,\psi\,|{\bf R}^{\prime}\}\, =
\,\delta({\bf R}_0-{\bf R}^{\prime})\,G_M({\bf P}_0)\,
\mathcal{F}^{(eq)}\{\phi |{\bf R}_0\}\,\,\,,
\end{array}
\]
\[
\mathcal{F}^{(eq)}\{\phi |{\bf R}_0\}\,\equiv \,1\,
+\sum_{n\,=1}^{\infty }\frac {\nu_{\,0}^n}{n!}\int^n_r
F_n^{(eq)}({\bf r}_1...\,{\bf r}_n|{\bf R}_0) \prod_{j\,=1}^n
\phi({\bf r}_j)\,\,,
\]
where functions $\,F_n^{(eq)}({\bf r}_1...\,{\bf r}_n|{\bf R}_0)\,$
are conditional thermodynamically equilibrium DF of the fluid,
corresponding to condition that BP is placed at point $\,{\bf
R}_0\,$, and\, $\,\mathcal{F}^{(eq)}\{\phi |{\bf R}_0\}\,$\, is
generating functional of these DF.

Notice, first, that $\,F_n^{(eq)}({\bf r}_1...\,{\bf r}_n|{\bf
R}_0)\,$ and thus $\,F_n(t=0,\,...\,)\,$ take account of all
correlations between BP and atoms, as well as atoms themselves, what
are caused by their interactions. Second, $\,F_n(t>0,\,...\,)\,$
treat essentially the same equilibrium state as at $\,t=0\,$.
Nevertheless time dependence of $\,F_n(t>0,\,...\,)\,$ does not copy
that of $\,V_0(t,{\bf R}_0,{\bf P}_0|{\bf R}^{\prime})\,$, i.e.
\[
\begin{array}{l}
F_n(t,{\bf R}_0, {\bf r}_1\,...\,{\bf r}_n,{\bf P}_0,{\bf
p}_1\,...\,{\bf p}_n|{\bf R}^{\prime})\,\neq\,\,
\,\,\,\,\,\,\,\,\,\,\,\,\,\,\,\,\,\,\,\,\,\,\,\,\,\,\,\,\,\\
\,\,\,\,\,\,\,\,\,\,\,\,\,\,\,\,\,\,\,
\,\,\,\,\,\,\,\,\,\,\,\,\,\,\,\,\,\,\,\,\,\,\,\,\,\, \neq\,V_0(t,{\bf
R}_0,{\bf P}_0|{\bf R}^{\prime})\,F_n^{(eq)}({\bf r}_1...\,{\bf
r}_n|{\bf R}_0)\prod_j G_m({\bf p}_j)\,
\end{array}
\]
Hence, at $\,t>0\,$ some extra correlations do arise between BP and
atoms. Statistical meaning of these correlations clears up if notice
that in absence of information on start position of BP, $\,{\bf
R}^{\prime}\,$, all joint DF of BP and atoms would be invariable.
This shows that in fact extra correlations connect total previous
BP's displacement, or path, $\,{\bf R}_0-{\bf R}^{\prime}\,$,
accumulated during time interval $\,(0,t)\,$, and current state of
the fluid at BP's surroundings. Therefore one may name them
``historical correlations''. Anyway, regardless of naming, they
should be attributed not to separate phase trajectories but to the
statistical ensemble of trajectories as the whole.

Let us introduce functions $\,V_n(t,{\bf R}_0,{\bf r}_1\,...\,{\bf
r}_n,{\bf P}_0,{\bf p}_1\,...\,{\bf p}_n|{\bf R}^{\prime})\,$ which
describe ``historical correlations'' in a pure form. By tradition,
differences of time-varying DF from equilibrium (or
quasi-equilibrium) DF are termed ``correlation functions'' (CF)
\cite{re,bog,bal}. We will apply this term also to the ``historical
correlations'' and define corresponding CF through their generating
functional as follows:
\[
\begin{array}{l}
\mathcal{F}\{t,{\bf R}_0,{\bf P}_0,\,\psi\,|{\bf R}^{\prime}\}\,\,=\,
\mathcal{F}^{(eq)}\{\phi |{\bf R}_0\}\,\,\mathcal{V}\{t,{\bf
R}_0,{\bf P}_0,\,\psi\,|{\bf R}^{\prime}\}\,\,\,,
\end{array}
\]
\begin{equation}
\begin{array}{l}
\mathcal{V}\{t,{\bf R}_0,{\bf P}_0,\,\psi\,|{\bf R}^{\prime}\}\,
\,=\, \,V_0(t,{\bf R}_0,{\bf P}_0|{\bf R}^{\prime})\,\,+
 \label{vf}
\end{array}
\end{equation}
\[
\,\,\,\,\,\,\,\,\,\,\,+\,\sum_{n\,=1}^{\infty } \frac
{\nu_{\,0}^n}{n!}\int^n_{r\times p} V_n(t,{\bf R}_0, {\bf
r}_1\,...\,{\bf r}_n,{\bf P}_0,{\bf p}_1\,...\,{\bf p}_n|{\bf
R}^{\prime})\prod_{j\,=1}^n \psi({\bf r}_j,-{\bf p}_j)\,\,
\]
with $\,\phi=\phi({\bf r})\,$ introduced in (\ref{cond}). At
$\,n=1\,$, in particular,
\begin{eqnarray}
F_1(t,{\bf R}_0,{\bf r}_1,{\bf P}_0,{\bf p}_1|{\bf
R}^{\prime})\,=\,\,\,\,\,\,\,\,\,\,\,\,\,\,\,\,\,\,\,\,
\,\,\,\,\,\,\,\,\,\,\,\,\,\,\,\,\,\,\,\,\,\,\,\,\,\,
\,\,\,\,\label{cf1} \\
=\, V_0(t,{\bf R}_0,{\bf P}_0|{\bf R}^{\prime})\, F_1^{(eq)}({\bf
r}_1|{\bf R}_0)\,G_m({\bf p}_1)\,+\, V_1(t,{\bf R}_0,{\bf r}_1,{\bf
P}_0,{\bf p}_1|{\bf R}^{\prime})\,\,\,,\nonumber
\end{eqnarray}
where function $\,V_1\,$ describes pair historical correlation of BP
with atoms.

According to definition (\ref{vf}) and above written initial
conditions for DF, all the CF vanish at start of BP's observation:\,
$\,V_n(t=0\,,...\,)=0\,$\, ($n> 0$). Next BP gradually gets dressed
in correlations with the fluid. In view of (\ref{unc}) and
(\ref{vf}),\, $\,V_n(t\,,...\,{\bf r}_k...\,)\rightarrow 0\,$\, at\,
$\,{\bf r}_k\rightarrow \infty\,$, that is the correlations always
vanish at infinity. Below we will see that their actual spatial
spread is very important characteristics of the Brownian motion under
consideration.

\section{Response of BP's path to fluid \\ perturbations and its
relations with CF}

Further, consider left side of identity (\ref{sim}) under choice
(\ref{ch}) rewriting it as
\begin{equation}
\begin{array}{l}
\int\int A({\bf q}(t),{\bf p}(t))\,B({\bf q},{\bf
p})\,\rho_{\,eq}({\bf q},{\bf p})\,d{\bf q}\,d{\bf p}\,=\,\int\int
\Psi ({\bf q},{\bf p})\,\rho_{\,eq}({\bf q},{\bf p})\,d{\bf q}\,d{\bf
p}\,\times\,\\ \label{ls}
\,\,\,\,\,\,\,\,\,\,\,\,\,\,\,\,\,\,\,\,\,\,\,\,\,\times\,\int\int
\delta({\bf R}(t)-{\bf R}^{\prime})\,\delta({\bf P}(0)-{\bf
P}_0)\,\rho_{\,in}({\bf q},{\bf p})\,d{\bf q}\,d{\bf p}\,\,,
\end{array}
\end{equation}
where\,\, $\,\Psi ({\bf q},{\bf p})=\,\Omega\,\delta({\bf R}-{\bf
R}_0)\prod_j\, [\,1+\psi({\bf r}_j,{\bf p}_j)\,]\,$\,\, and
\[
\begin{array}{c}
\rho_{\,in}({\bf q},{\bf p})\,=\, \Psi ({\bf q},{\bf
p})\,\rho_{\,eq}({\bf q},{\bf p})\,[\,\int\int \Psi ({\bf
q}^{\prime},{\bf p}^{\prime})\,\rho_{\,eq}({\bf q}^{\prime},{\bf
p}^{\prime})\,d{\bf q}^{\prime}\,d{\bf p}^{\prime}]^{-\,1}\,
\end{array}
\]
Evidently, last integral in (\ref{ls}) represents averaging over new
statistical ensemble of initial conditions, with new probabilistic
measure $\,\rho_{\,in}({\bf q},{\bf p})\,$. The latter is
thermodynamically non-equilibrium measure which however would be a
canonical equilibrium one in presence of imaginary
(momenta-dependent) external potential $\,U({\bf q},{\bf
p})=-T\,\ln{[1+\psi({\bf q},{\bf p})\,]}\,$. Hence,
\begin{equation}
\begin{array}{c}
\int\int \delta({\bf R}(t)-{\bf R}^{\prime})\,\delta({\bf P}(0)-{\bf
P}_0)\,\rho_{\,in}({\bf q},{\bf p})\,d{\bf q}\,d{\bf
p}\,=\\
=\,V\{t,{\bf R}^{\,\prime}|\psi ,{\bf R}_0 ,{\bf P}_0\}\,G_M({\bf
P}_0)\,\,\,,\label{ls1}
\end{array}
\end{equation}
where $\,V\{t,{\bf R}^{\,\prime}|\psi ,{\bf R}_0 ,{\bf P}_0\}\,$ is
conditional probability density of finding BP at $\,t>0\,$ at point
$\,{\bf R}^{\,\prime}\,$ under conditions that initially, at
$\,t=0\,$, it was located at point $\,{\bf R}_0\,$ with momentum
$\,{\bf P}_0\,$ while the fluid was in such non-equilibrium spatially
non-uniform state what turns into equilibrium under the external
potential $\,U({\bf q},{\bf p})\,$. Noticing, besides, that
\[
\begin{array}{l}
\int\int \Psi ({\bf q},{\bf p})\,\rho_{\,eq}({\bf q},{\bf p})\,d{\bf
q}\,d{\bf p}\,=\, \mathcal{F}^{(eq)}\{\phi |{\bf R}_0\}\,\,\,,
\end{array}
\]
we can transform equality (\ref{ls}) to
\begin{equation}
\begin{array}{l}
\int\int A({\bf q}(t),{\bf p}(t))\,B({\bf q},{\bf
p})\,\rho_{\,eq}({\bf q},{\bf p})\,d{\bf q}\,d{\bf p}\,=
\,\,\,\,\,\,\,\,\,\,\,\,\,\,\,\\\,\,\,\,\, \,\,\,\,\, =\,V\{t,{\bf
R}^{\,\prime}|\psi ,{\bf R}_0 ,{\bf P}_0\}\,\,G_M({\bf
P}_0)\,\,\mathcal{F}^{(eq)}\{\phi |{\bf R}_0\}\, \label{ls2}
\end{array}
\end{equation}

Instead of $\,\psi({\bf q},{\bf p})\,$ or $\,U({\bf q},{\bf p})\,$,
the non-equilibrium ensemble determined by $\,\rho_{\,in}({\bf
q},{\bf p})\,$ can be characterized in terms of corresponding initial
mean densities of atoms in the $\,\mu$-space, $\,\mu\{{\bf r},{\bf
p}|\,\psi ,{\bf R}_0 \}\,$, and coordinate space, $\,\nu\{{\bf
r}|\,\phi ,{\bf R}_0 \}=\int \mu\{{\bf r},{\bf p}|\,\psi ,{\bf R}_0
\}\,d{\bf p}\,$. More precisely, that are conditional densities at
given BP's position. In other words, $\,\mu\{{\bf r},{\bf p}|\,\psi
,{\bf R}_0 \}\,$ is initial one-particle DF of the fluid. It is easy
to verify that
\begin{equation}
\mu\{{\bf r},{\bf p}|\,\psi ,{\bf R}_0 \}\, =\,[\,1+\psi({\bf r},{\bf
p})\,]\,\,G_m({\bf p})\,\,\frac {\delta \ln\mathcal{F}^{(eq)}\{\phi
|{\bf R}_0 \}}{\delta\phi({\bf r})} \,\label{n}
\end{equation}

After all, combining formulas (\ref{sim}), (\ref{gf}), (\ref{vf}) and
(\ref{ls2}), we come to our main formally exact relation:
\begin{eqnarray}
V\{t,{\bf R}^{\,\prime}|\psi ,{\bf R}_0 ,{\bf P}_0\}\,\,G_M({\bf
P}_0)\,=\,V_0(t,{\bf R}_0,-{\bf P}_0|{\bf R}^{\prime})\,+
\,\,\,\,\,\,\,\,\,\,\,\,\,\,\label{r}\\
+\,\sum_{n\,=1}^{\infty } \frac {\nu_{\,0}^n}{n!}\int^n_{r\times p}
V_n(t,{\bf R}_0,{\bf r}_1...\,{\bf r}_n,-{\bf P}_0,{\bf
p}_1\,...\,{\bf p}_n|{\bf R}^{\,\prime}) \prod_{j\,=1}^n \psi({\bf
r}_j,-{\bf p}_j)\,\nonumber
\end{eqnarray}
It connects, from left hand, probability distribution of BP's path in
initially non-equilibrium nonuniform fluid and, from right hand,
similar distribution, along with generating functional of
correlations between the path and current BP's environment, for
equilibrium uniform fluid. At $\,\psi({\bf r},{\bf p})=0\,$ this is
merely time symmetry relation for equilibrium random walk:
\begin{equation}
\begin{array}{l}
V_0(t,{\bf R}^{\,\prime}|{\bf R}_0 ,{\bf P}_0)\,\,G_M({\bf
P}_0)\,=\,V_0(t,{\bf R}_0,-{\bf P}_0|{\bf R}^{\prime})\,\,\label{r00}
\end{array}
\end{equation}

Of course, all the DF and CF are dependent on the mean gas density
$\,\nu_{\,0}\,$. But for brevity in relation (\ref{r}), as well as
before it and almost everywhere below, corresponding argument is not
written out.

\section{Virial expansion}
Let us choose $\,\psi({\bf r},{\bf p})\,$ independent on momentum,
$\,\psi({\bf r},{\bf p})=\phi({\bf r})\,$, with $\,\phi({\bf r})\,$
being constant, $\,\phi({\bf r})=\phi =\,$const\,, inside some sphere
$\,|{\bf r}-{\bf R}_0|< \xi\,$ and vanishing outside it in a suitable
way. Then factual perturbation of spatial uniformity of the fluid and
thus of its equilibrium initially takes place at $\,|{\bf r}-{\bf
R}_0|> \xi\,$ only.

Condition (\ref{cond}) requires from the radius $\,\xi\,$ to be
finite. Nevertheless we can make it as large as we want. Therefore,
at any given finite time $\,t_0\,$ we can find such finite $\,\xi\,$
that at $\,t<t_0\,$ the whole BP's path on left side of (\ref{r}) for
sure realizes in uniform equilibrium fluid, although with a density
$\,\nu \,$ which differs from the density $\,\nu_{0} \,$ on the
right-hand side. For instance, already such value as $\,\xi = k\,v_s
t_0\,$, where $\,v_s\,$ is speed of sound in our fluid and $\,k>2\,$,
seems to be sufficient.

Under these assumptions, the functional $\,V\{t,{\bf R}^{\prime}|\phi
,{\bf R}_0,{\bf P}_0 \}\,$ simplifies to mere function, and for
$\,t<t_0\,$ we can write
\begin{equation}
\begin{array}{l}
\int V\{t,{\bf R}^{\prime}|\phi ,{\bf R}_0,{\bf P}_0 \}\,G_M({\bf
P}_0)\,d{\bf P}_0\,=\,V_0(t,{\bf R}^{\prime}-{\bf R}_0\,;\,\nu
)\,\,\,,\label{int}
\end{array}
\end{equation}
where third density argument, $\,\nu \,$, of the BP's path
distribution is introduced, so that $\,V_0(t,{\bf R}_0-{\bf
R}^{\prime}\,;\,\nu_0\,)\,=\,V_0(t,{\bf R}_0-{\bf R}^{\prime})\,$. On
right-hand side of (\ref{r}), under same conditions, none correlation
between BP and fluid may propagate up to distances $\,|{\bf r}-{\bf
R}_0|> \xi\,$. Therefore in all integrals $\,\phi({\bf r})\,$ can be
replaced by the constant.

These reasonings show that relation (\ref{r}) tolerates extension to
$\,\psi =\phi =\,$const\,. At that after use of (\ref{r00}) and
(\ref{int}) it transforms to
\begin{equation}
V_0(t,\Delta{\bf R}\,;\,\nu(\nu_{\,0},\phi ))\,=\,V_0(t,\Delta{\bf
R}\,;\,\nu_{\,0}) \,+\, \sum_{n\,=\,1}^{\infty } \,\frac {\phi^n}{n!}
\,V_n(t,\Delta{\bf R}\,;\,\nu_{\,0})\,\,\label{ve}
\end{equation}
Here $\,\Delta{\bf R}\equiv {\bf R}_0-{\bf R}^{\prime}\,$, functions
$\,V_n\,$ with $\,n>0\,$ are defined by
\begin{equation}
V_n(t,\Delta{\bf R}\,;\nu_{\,0})=\,\nu_{\,0}^n\,\int^n_{r\times p}
\int V_n(t,{\bf R}_0,{\bf r}_1...\, {\bf r}_n,{\bf P}_0,{\bf
p}_1...\, {\bf p}_n\,|\,{\bf R}^{\prime})\,d{\bf P}_0\,\,
\label{clim0}
\end{equation}
and\, $\,\nu(\nu_{\,0},\phi )\,$ is the left fluid density $\,\nu\,$
as a function $\,\nu_{\,0}\,$ and $\,\phi\,$. Evidently, that is
$\,\nu\{{\bf r}|\phi ,{\bf R}_0 \}\,$ taken far from BP, at $\,|{\bf
r}-{\bf R}_0|\gg r_b\,$, where $\,r_b\,$ is characteristic radius of
BP-atom interaction potential $\,U_{ab}(r)\,$. Due to (\ref{n}),
\begin{equation}
\begin{array}{l}
\nu =\nu(\nu_{\,0}\,,\phi )\,=\,\nu_{\,0}\,(1+\phi
)\,\{\,1+\nu_{\,0}\phi \int [\,F^{(eq)}_2({\bf r})-1]\,d{\bf
r}\,\,+\,\,\,\, \label{n1}
\end{array}
\end{equation}
\[
\,+\,\frac {\nu_{\,0}^2\phi^2}{2} \int_r^{\,2} [\,F^{(eq)}_3({\bf
r}_1,{\bf r}_2,0)-F^{(eq)}_2({\bf r}_1)-F^{(eq)}_2({\bf
r}_2)-F^{(eq)}_2({\bf r}_1-{\bf r}_2)+2\,]\, +...\,\}
\]
with $\,F^{(eq)}_2({\bf r})\,$ and $\,F^{(eq)}_3({\bf r}_1,{\bf
r}_2,{\bf r}_3)\,$ being standard equilibrium pair and triple DF at
density $\,\nu_{\,0}\,$, and $\,\int _r^2\,$ symbolizes integration
over $\,{\bf r}_1\,$ and $\,{\bf r}_2\,$.

Formula (\ref{ve}) may be rated as virial expansion of the BP's path
probability distribution. But it significantly differs from the
well-known virial expansions of thermodynamical quantities \cite{ll1}
or kinetic coefficients \cite{ll2}: the latter expand over absolute
value of density $\,\nu\,$ while (\ref{ve}) in fact over difference
$\,\nu/\nu_0 -1\,$ or logarithm $\,\ln{(\nu/\nu_0)}\,$.

For further, the first-order term of the virial expansion is most
important. Differentiation of (\ref{ve}) with respect to $\,\phi\,$ ,
at $\,\phi =0\,$, together with (\ref{clim0}) and (\ref{n1}) yields
\begin{eqnarray}
\widetilde{\nu }_0\,\frac {\partial V_0(t,\Delta {\bf
R}\,;\,\nu_{\,0})}{\partial \nu_{\,0}}\, = \,V_1(t,\Delta {\bf
R}\,;\,\nu_{\,0})\,=\,\,\,\,\,\,\,\,\,\,\,\,\,\,\,\,\,\,\,\,\,
\,\,\,\,\,\,\,\,\,\,\,\,\,\,\,\,\,\,\,\,\label{r21}
\\ \,\,\,\,\,\,\,\,\,\,\,\,\,\,\,\,\,\,\,\,\,\,\,
\,\,\,\,\,\,\,\,\,\,\,
=\,\nu_{0}\int\int\int V_1(t,{\bf R}_0,{\bf P}_0,{\bf r},{\bf
p}\,|{\bf R}^{\prime})\,\,d{\bf r}\, \,d{\bf P}_0\,d{\bf p}\,\,\,,
\nonumber
\end{eqnarray}
\begin{eqnarray}
\widetilde{\nu }_0\,\equiv \,\left [\frac {\partial \nu
(\nu_{\,0},\phi )}{\partial \phi }\right ]_{\phi\, =0} =\,\nu_0
+\nu_0^2\,\int [F_2^{(eq)}({\bf r})-1\,]\,d{\bf
r}=\,\nu_{\,0}\,T\left(\frac {\partial \nu_{\,0}}{\partial
P}\right)_T\,\,\, \label{n2}
\end{eqnarray}
The last equality, where $\,P\,$ denotes pressure, is known from the
classical statistical thermodynamics \cite{ll1}.

\section{Restrictions on BP's path distribution}
Relations (\ref{ve}) with (\ref{clim0}) and (\ref{r21}) with
(\ref{n2}) prove that historical correlations between BP and fluid
really exist. Let us show that they imply essential restrictions on
possible shape of the distribution $\,V_0(t,\Delta {\bf
R}\,;\nu_0)\,$.

Consider the pair correlation $\,V_1(t,{\bf R}_0,{\bf P}_0,{\bf
r},{\bf p}\,|{\bf R}^{\prime})\,$. Let from now $\,\Omega \,$ denotes
a finite region in the $\,({\bf r}-{\bf R}_0)$-space and at once
volume of this region. Introduce\, $\,\Omega (\delta)=\Omega(t,\Delta
{\bf R},{\bf P}_0,{\bf p},\delta)\,$\, as minimum of all such regions
$\,\Omega \,$ what satisfy
\begin{equation}
\begin{array}{c}
\left |\,\int_{\Omega} V_1(t,{\bf R}_0,{\bf r},{\bf P}_0,{\bf p}|{\bf
R}^{\prime}\,)\,d{\bf r}\,-\,\int V_1(t,{\bf R}_0,{\bf r},{\bf
P}_0,{\bf p}|{\bf R}^{\prime}\,)\,d{\bf r}\,\right |\,<\,
\,\,\,\,\,\,\,\,\,\\ \,\,\,\,<\,\delta\,\left |\,\int V_1(t,{\bf
R}_0,{\bf r},{\bf P}_0,{\bf p}|{\bf R}^{\prime}\,)\,d{\bf r}\,\right
|\,\label{in0}
\end{array}
\end{equation}
with some fixed $\,0< \delta < 1\,$. Thus
$\,\Omega(\delta)=\Omega(t,\Delta {\bf R},{\bf P}_0,{\bf
p},\delta)\,$ is minimal region containing at least $\,100\,(1-\delta
)\,$ percents of the total pair correlation.

From the other hand, integrate identity (\ref{cf1}) over $\,{\bf r}_1
\in \Omega(\delta ) \,$. Since (\ref{cf1}) represents a probability
density, result must be non-negative:
\[
\begin{array}{c}
V_0(t,{\bf R}_0,{\bf P}_0|{\bf R}^{\prime})\,G_m({\bf
p})\,\int_{\,\Omega(\delta )} F_1^{(eq)}({\bf r}|{\bf R}_0)\,d{\bf
r}\,+ \,\,\,\,\,\,\,\,\,\,\,\,\,\,\,\,\,\,\,\,\,\,\,\,\,\,\,\,
\,\,\,\,\,\,\,\,\,\,\,\,\,\,\,\,\,\,\,\,\,\\ \,\,\,\,\,\,\,\,\,
\,\,\,\,\,\,\,\,\,\,\,\,\,\,\,\,\,\,\,\,\,\,\,\,\,\,\,\,\,\,
+\,\int_{\,\Omega(\delta )} V_1(t,{\bf R}_0,{\bf r},{\bf P}_0,{\bf
p}|{\bf R}^{\prime}) \,d{\bf r}\,\geq\,0\,
\end{array}
\]
This inequality together with (\ref{in0}) yields more interesting
one:
\begin{equation}
\begin{array}{c}
V_0(t,{\bf R}_0,{\bf P}_0|{\bf R}^{\prime})\,G_m({\bf
p})\,\int_{\,\Omega(\delta)} F_1^{(eq)}({\bf r}|{\bf R}_0)\,d{\bf
r}\, +\,\,\,\,\,\,\,\,\,\,\,\,\,\,\,\,\,\,\,\,\,\,\,\,\,\,
\,\,\,\,\,\,\,\,\,\,\,\,\,\,\,\,\,\,\,\,\, \label{in}
\\\,\,\,\,\,\,\,\,\,\,\,\,\,\,\,\,\,\,\,\,\,\,\,\,\,\,\,\,\,
\,\,\,\,\,\,\,\,\,\,\,\,\, +\,(1-\delta\,)\int V_1(t,{\bf R}_0,{\bf
r},{\bf P}_0,{\bf p}|{\bf R}^{\prime}) \,d{\bf r}\,\,\geq\,\,0\,
\end{array}
\end{equation}
Next perform here integration over momenta, apply virial relation
(\ref{r21}) and then divide all by $\,(\,1-\delta )\,$ and multiply
by $\,\nu_0\,$. Thus we obtain
\begin{eqnarray}
\nu_{0}\,\,\overline{\Omega}(t,\Delta {\bf
R},\,\delta)\,\,V_0(t,\Delta {\bf
R}\,;\,\nu_{\,0})\,+\,\,\widetilde{\nu }_0\,\frac {\partial
V_0(t,\Delta {\bf R}\,;\,\nu_{\,0})}{\partial \nu_{\,0}}\,\, \geq\,0
\label{in5}
\end{eqnarray}
Here $\,\overline{\Omega}(t,\Delta {\bf R},\,\delta)\,$ is
characteristic pair correlation volume defined by
\begin{equation}
\begin{array}{c}
\overline{\Omega}(t,\Delta {\bf R},\,\delta)\,=\, (\,1-\delta
)^{-\,1}\,\int \int G_M(t,{\bf P}_0|\Delta {\bf R})\,G_m({\bf
p})\,\times\,\,\,\,\,\,\,\,\,\,\,\,\,\,\,\,\,\,\,
\,\,\,\,\,\,\,\,\,\,\,\,\,\,\,\,\,\,\,\,\,\, \label{av1}
\\\,\,\,\,\,\,\,\,\,\,\,\,\,\,
\,\,\,\,\,\,\,\,\,\,\,\,\,\,\,\,\,\,\,\,\,
\,\,\,\,\,\,\,\,\,\,\times\,\left [\,\int_{\,\Omega(\,t,\,\Delta {\bf
R},\,{\bf P}_0,\,{\bf p},\,\delta)}\,F_1^{(eq)}({\bf r}|{\bf
R}_0)\,d{\bf r}\,\right ]\,d{\bf p}\,d{\bf P}_0 \,
\end{array}
\end{equation}
with\,\,\, $\,G_M(t,{\bf P}_0|\Delta {\bf R})\,\equiv\, V_0(t,{\bf
R}_0,{\bf P}_0|{\bf R}^{\prime})/V_0(t,\Delta {\bf R})\,$\,\, being
conditional distribution of momentum of BP under given value of its
path.

Inequality (\ref{in5}) presents us a restriction on possible negative
values of derivative\, $\,\partial \ln V_0(t,\Delta {\bf
R}\,;\,\nu_{\,0})/\partial \ln \nu_{\,0}\,$. The smaller is
$\,\overline{\Omega}(t,\Delta {\bf R},\,\delta)\,$ the stronger is
this restriction. Therefore $\,\overline{\Omega}(t,\Delta {\bf
R},\,\delta)\,$ can be replaced by
\[
\overline{\Omega}(t,\Delta {\bf R})\,=\,\min_{0<\,\delta<\,1}
\overline{\Omega}(t,\Delta {\bf R},\,\delta)\,
\]

Now let us suppose that at sufficiently large temporal and spatial
scales, at $\,t\gg \tau \,$ and $\,|\Delta {\bf R}|\gg \Lambda\,$,
with $\,\tau \,$ and $\,\Lambda =v_T\tau\,$ being BP's mean
free-flight time and mean free path, respectively (and $\,v_T\sim
\sqrt{T/M}\,$ its thermal velocity), the BP's path distribution tends
to the Gaussian distribution:
\begin{equation}
\begin{array}{c}
V_0(t,\Delta{\bf R})\,\rightarrow\,V_G(t,\Delta{\bf R})\,=\,(4\pi
Dt)^{-\,3/2}\,\exp{(-\Delta {\bf R}^2/4Dt)}\,\, \label{vg}
\end{array}
\end{equation}
Then inequality (\ref{in5}), with use of (\ref{n2}), takes the form
\begin{eqnarray}
\left [\,\nu_{0}\,\overline{\Omega}(t,\Delta {\bf R})\,+\,\frac
{\widetilde{\nu }_{\,0}}{D}\,\frac {\partial D}{\partial
\nu_{\,0}}\,\left (\frac {\Delta{\bf R}^2}{4Dt}-\frac 32 \right
)\,\right ]V_G(t,\Delta{\bf R})\,\geq\,0\,\,\label{ga}
\end{eqnarray}
In natural situation when BP's diffusivity is a decreasing function
of the density, $\,\partial D/\partial \nu_{\,0}<0\,$ (e.g. when our
fluid is a gas), the second addend in square bracket becomes negative
at $\,z\equiv \Delta{\bf R}^2/4Dt\,>\,3/2\,$. As the consequence,
inequality (\ref{ga}) can be satisfied if and only if at large values
of $\,z\,$ the volume $\,\overline{\Omega}(t,\Delta {\bf R})\,$
infinitely grows, at least proportionally to $\,z\,$.

Thus, we came to intriguing paradox or theorem. If asymptotical
statistics of the BP's path is Gaussian then characteristic volume
occupied by pair correlation of BP with atoms is unbounded above. In
opposite, if the correlation volume is bounded above,
$\,\overline{\Omega}(t,\Delta {\bf R})<\overline{\Omega}_{max}\,$
with $\,\overline{\Omega}_{max}\,$ being independent on $\,t\,$ and
$\,\Delta {\bf R}\,$, then asymptotic of $\,V_0(t,\Delta{\bf R})\,$
must be essentially non-Gaussian (see below).

What is closer to reality? Further, we want to show that the second
statement conforms to truth at least if our fluid is a dilute gas.

\section{Pair correlation volume in dilute gas \\
and the Boltzmann-Grad limit}

For our purpose, to exclude non-principal complications, the
Boltzmann-Grad limit (BGL) is very useful, when
$\,\nu_{0}\rightarrow\infty\,$ while $\,r_b\sim r_a \rightarrow 0\,$
($\,r_a\,$ is radius of short-range repulsive atom-atom interaction)
in such way that gas non-ideality parameters $\,4\pi r_a^3\nu_0/3\,$
and $\,4\pi r_b^3\nu_0/3\,$ vanish but mean free paths of BP,
$\,\Lambda =(\pi r_b^2\nu_0 )^{-1}\,$, and atoms, $\,\lambda =(\pi
r_b^2\nu_0 )^{-1}\,$, stay fixed.

It is easy to justify that under BGL expression (\ref{n}) effectively
simplifies:\,  $\,\mu\{{\bf r},{\bf p}|\psi,{\bf
R}_0\}/\nu_{0}\,\rightarrow\, [\,1+\psi({\bf r},{\bf p})\,]\,G_m({\bf
p})\,$\,,\, function $\,F_1^{(eq)}({\bf r}|{\bf R}_0)\,$ in
(\ref{av1}) can be replaced by unit, and inequality (\ref{in5})
yields
\begin{eqnarray}
\nu_{0}\,\,\overline{\Omega}(t,\Delta {\bf R})\,\,V_0(t,\Delta {\bf
R}\,;\,\nu_{\,0})\,+\,\,\nu_0\,\frac {\partial V_0(t,\Delta {\bf
R}\,;\,\nu_{\,0})}{\partial \nu_{\,0}}\,\, \geq\,0\,\,\,,\label{in6}
\end{eqnarray}
\[
\overline{\Omega}(t,\Delta {\bf R})=\int \int G_M(t,{\bf P}_0|\Delta
{\bf R})\,G_m({\bf p})\, \min_{0<\,\delta<\,1}\frac {\Omega(t,\Delta
{\bf R},{\bf P}_0,{\bf p},\delta)}{1-\delta }\,\,d{\bf p}\,d{\bf P}_0
\,
\]

Let us discuss the pair correlation volume, basing on the first-order
term of the expansion (\ref{r}):
\begin{equation}
G_M({\bf P}_0)\left [\frac {\delta V\{t,{\bf R}^{\,\prime}|\psi ,{\bf
R}_0 ,{\bf P}_0\}}{\delta \psi({\bf r},{\bf p})}\right ]_{\psi\,=\,0}
=\nu_{\,0}V_1(t,{\bf R}_0,{\bf r},-{\bf P}_0,-{\bf p}\,|{\bf
R}^{\,\prime}\,) \label{fo11}
\end{equation}
At that, we will take in mind also a few things about pair
correlations and pair CF known from classical theory of dilute gases
\cite{re,bog,uf,bal} and, besides, those distinctive property of the
Boltzmann-Grad gas that any two of its particles deal with different
non-intercrossing collections of other particles, except cases when
direct encounter of the two takes place.

Notice that left side of (\ref{fo11}) reflects reaction of
probability distribution of BP's path, $\,{\bf R}^{\,\prime}-{\bf
R}_0\,$, to lodging, at $\,t=0\,$, in point $\,{\bf r}\,$\, one extra
atom with momentum $\,{\bf p}\,$. In view of the stated property,
this reaction is nonzero at such initial relative disposition of BP
and extra atom only which ensures their direct collision, either soon
forthcoming or just happening one. For that, two conditions should be
satisfied. Firstly, vector $\,{\bf r}-{\bf R}_0\,$ belongs to
so-called ``collision cylinder'' which has radius $\,\approx r_b\,$
and is oriented along relative velocity of the particles, $\,{\bf
v}-{\bf V}_0={\bf p}/m-{\bf P}_0/M\,$. Secondly, $\,|{\bf r}-{\bf
R}_0|\,$ is not much greater than $\,\Lambda\,$ or $\,\lambda\,$.
Otherwise collisions of either BP or given extra atom with other
atoms (the rest of gas) will prevent the desired collision. These
remarks, in the framework of (\ref{fo11}), make it obvious that
volume occupied by the pair correlation can be estimated as
\[
\begin{array}{l}
\Omega(t,\Delta {\bf R},{\bf P}_0,{\bf p},\delta)\,\sim\,\pi
r_b^2\Lambda\,=\,\nu_0^{-1}\,\,\,\,\,\,\,\,\,\,\,
\end{array}
\]
(assuming for simplicity that $\,\lambda\sim\Lambda\,$), i.e. it is
comparable with volume displayed per one atom of the gas.

The first condition agrees with the classical theory: the non-zero
reaction on the left in (\ref{fo11}) corresponds to non-zero
correlation on the right well-known at least for post-collision
dispositions. However, the second condition does not agree: a spread
of this correlation along the collision cylinder in classical theory
remains indefinite. Formally, infinite. The matter is that this
theory neglects destruction of pair correlation by collisions of the
pair with other particles. This leads to infinite volume of pair
correlation, which is highly doubtful result.

At the same time, the theory \cite{re,bog,uf,bal} says that magnitude
of pair CF, $\,V_1(t,{\bf R}_0,{\bf r},{\bf P}_0,{\bf p}|{\bf
R}^{\prime})\,$, inside collision cylinder generally is comparable
with product of one-particle DF (first right-hand term in
(\ref{cf1})). This is very important statement, and it looks, at
least for $\,|{\bf r}-{\bf R}_0|<\Lambda\,$, as quite doubtless
result. It follows also from the first of the BBGKY equations under
remark that first term of (\ref{cf1}) does not contribute to the
``collision integral'' (see \cite{i1,i2,p1,p2}). In view of the exact
virial relation (\ref{r21}) this result clearly cancels the previous
one. Moreover, confirms that in fact volume of pair correlation has
an order of $\,\nu_0^{-1}\,$.

Importantly, this estimate manifests no visible dependence on either
$\,t\,$ and $\,\Delta {\bf R}\,$ or the momenta. Therefore, taking
into account that at $\,t\gg\tau\,$ certainly $\,G_M(t,{\bf
P}_0|\Delta {\bf R})\approx G_M({\bf P}_0)\,$ (see
\cite{i1,i2,p1,p2}), we can write
\begin{eqnarray}
\overline{\Omega}(t,\Delta {\bf
R})\,\approx\,\,\min_{0<\,\delta<\,1}\frac {\Omega(\delta)}{1-\delta
}\,\,\leq\,\overline{\Omega}_{max}\,\sim\,\nu_0^{-1}\,\label{av2}
\end{eqnarray}
with $\,\Omega(\delta)\,$ representing $\,\Omega(t,\Delta {\bf
R},{\bf P}_0,{\bf p},\delta)\,$ at typical thermal momenta (see its
definition in previous section). As the consequence, inequality
(\ref{in6}) can be replaced by
\begin{eqnarray}
c_1\,V_0(t,\Delta {\bf R}\,;\,\nu_{\,0})\,+\,\,\nu_0\,\frac {\partial
V_0(t,\Delta {\bf R}\,;\,\nu_{\,0})}{\partial \nu_{\,0}}\,\,
\geq\,0\,\,\,,\label{in7}
\end{eqnarray}
where\, $\,c_1\,=\,\nu_{\,0}\,\overline{\Omega}_{max}\,\sim\,1\,$\,
is a constant of order of unit.

\section{Probability distribution of molecular \\ random walk in dilute
gas has long tails}

The Gaussian asymptotic (\ref{vg}) is incompatible with inequality
(\ref{in7}). To see what is allowed instead of (\ref{vg}), it is
natural to suppose that at $\,t\gg\tau\,$ and $\,|\Delta {\bf
R}|\gg\Lambda\,$ the BP's path distribution $\,V_0(t,\Delta {\bf
R}\,;\,\nu_{\,0})\,$ is characterized, similarly to $\,V_G(t,\Delta
{\bf R})\,$, by a single parameter, that is diffusivity. Then
\begin{equation}
\begin{array}{c}
 V_0(t,\Delta{\bf R}\,;\,\nu_{\,0}) \rightarrow
(4Dt)^{-3/2}\,\Psi(\Delta {\bf R}^2/4Dt)\,\,\label{as0}
\end{array}
\end{equation}
Conditions\, $\,\int V_0\,d\Delta {\bf R}\,=\,1\,$ and $\,\int \Delta
{\bf R}^2\,V_0\,d\Delta {\bf R}\,=\,6Dt\,$\, (which in fact serves as
quantitative definition of the diffusivity) imply that
\[
\begin{array}{c}
\int \Psi({\bf a}^2)\,d{\bf a}\,=1\,\,\,,\,\,\,\,\,\,\int {\bf
a}^2\Psi({\bf a}^2)\,d{\bf a}\,=3/2\,\,\,
\end{array}
\]
Since in a dilute (Boltzmann-Grad) gas diffusivity is inversely
proportional to gas density, $\,D=v_T\Lambda\propto\nu_0^{-1}\,$,
that is $\,\nu_{\,0}\,\partial D/\partial \nu_{\,0} =-D\,$,
inequality (\ref{in7}) yields
\begin{equation}
\alpha\,\Psi(z) +z\,\frac {d\,\Psi(z)}{d\,z} \,\geq
\,0\,\,\,\,\,,\,\,\,\,\,\,\,\,\alpha \,\equiv\,\frac 32\,+\,c_1
\,\,\label{as1}
\end{equation}
Hence, function $\,\Psi(z)\,$ must have power-law long tail: \,
$\,\Psi(z)\propto\, z^{-\,\beta }\,$\, at $\,z\rightarrow\infty\,$\,,
where $\,\beta\leq \alpha\,$\,. At that, the second condition
requires $\,\alpha >5/2\,$.

One may rightly object that such asymptotic is nonphysical since has
infinite higher-order statistical moments of $\,\Delta{\bf R}\,$ and
makes probable enormously fast displacement of BP. But we can easy
correct all this if recall that in addition to diffusivity the BP's
motion has one more significant parameter, namely, BP's thermal
velocity $\,v_T\,$, which is independent on $\,\nu_{\,0}\,$ and not
touched upon by (\ref{in7}). Hence, a true asymptotic has the form
\begin{equation}
V_0(t,\Delta {\bf R})\rightarrow \frac {1}{(4Dt)^{3/2}}\, \Phi\left
[\frac {\Delta{\bf R}^2}{4Dt}\,,\frac {|\Delta {\bf R}|}{v_T t}\right
]\approx \frac {1}{(4Dt)^{3/2}}\,\Psi\left (\frac {\Delta{\bf
R}^2}{4Dt}\right)\Theta \left [\frac {|\Delta {\bf R}|}{v_Tt}\right
]\,\,\label{as2}
\end{equation}
Here function $\,\Theta(y)\approx 1\,$ at $\,y\ll 1\,$ and fast
enough tends to zero at $\,y>1\,$ thus sharply cutting long tails of
$\,V_0(t,\Delta{\bf R})\,$ when\, $\,|\Delta {\bf R}|>v_Tt\,$\,.

\section{Comparison with earlier results}

To get more details of $\,V_0(t,\Delta{\bf R})$'s behavior one has to
supplement virial relations by additional tools, in principle, by
corresponding Bogolyubov-Born-Green-Kirkwood-Yvon (BBGKY) equations.
Anyway, remarkably, our present results are in full qualitative
agreement with what was obtained in \cite{i1} (or see \cite{i2}) and
in \cite{p1} from approximative analysis of BBGKY equations for the
Boltzmann-Grad gas (with marked gas atom in the role of BP). An
expression for the BP's path distribution, $\,V_0(t,\Delta{\bf R})$,
found in$\,^2\,$ has just the form (\ref{as2}) with\,
$\,\Psi(z)\,=\,\Gamma (7/2)\, (1+z)^{-\,7/2}\,$,\, that is $\,\alpha
=7/2\,$, which corresponds to $\,c_1=\nu_0\overline{\Omega
}_{max}=2\,$. At that, the cut-off function $\,\Theta\,$ starts to
work already from fourth-order statistical moment:
\begin{equation}
\langle\, \Delta{\bf R}^4(t) \,\rangle \,=\,\int \Delta{\bf
R}^4\,V_0(t,\Delta {\bf R})\,d\Delta{\bf
R}\,\rightarrow\,3\,\langle\, \Delta{\bf R}^2(t) \,\rangle
^2\,\ln{\frac {t}{\tau }} \,\label{fm}
\end{equation}
with $\,\langle \Delta{\bf R}^2(t) \rangle =6Dt\,$. This agrees with
results of \cite{i1} found in different approach when only four
lowest moments were under consideration.

Such behavior of the fourth-order statistical moment as presented by
(\ref{fm}) can be phenomenologically interpreted as manifestation of
scaleless 1/f fluctuations in diffusivity (and mobility) of molecular
Brownian particle. Thus, the present work sustains the theory of
microscopic Brownian motion and related 1/f-noise in (infinitely)
many-particle systems which was first suggested in works
\cite{bk1,bk3}. Heuristic explanations of statistical nature of these
fluctuations can be found also in \cite{i1,i2,p1,pro,last}.

\section{Conclusion}

We derived exact relations which connect probability distribution of
path of molecular Brownian particle (BP) in a perturbed fluid and
specific ``historical'' statistical correlations between the BP's
path and molecules of equilibrium fluid. The simplest of this
relations produced well rigorous differential inequality which says
that Gaussian asymptotic of the path distribution is incompatible
with finiteness of spatial spread of the pair historical correlation,
even in dilute gas under the Boltzmann-Grad limit. Then we argued
that indeed the spread is bounded above. Consequently, the path
distribution has essentially non-Gaussian form, with power-law long
tails (lasting up to $\,|\Delta {\bf R}|/t\,$\, or order of BP's
thermal velocity).

These results mean that BP's path, even in the Boltzmann-Grad gas
(moreover, first of all in it), can not be disjointed into some parts
what are statistically independent in the sense of the probability
theory. In other words, real microscopic dynamics of (infinitely)
many-particle system produces much more rich randomness than poor
probability-theoretical ``dice tossing'' can do. That is occasion to
recall the A.\,Einstein's words: ``God does not play dice''.

I acknowledge my colleagues Dr.\,\,Yu.\,Medvedev and
Dr.\,\,I.\,Krasnyuk for interesting and useful discussions.

\end{document}